# A POSSIBLE HARAPPAN ASTRONOMICAL OBSERVATORY AT DHOLAVIRA

**Mayank Vahia**

*Tata Institute of Fundamental Research, Homi Bhaba Road,
Colaba, Mumbai 400 005, India, and Manipal Advanced Research Group,
Manipal University, Manipal, Karnataka 576 104, India.*
Email: vahia@tifr.res.in

*and*

**Srikumar M. Menon**

*Manipal School of Architecture and Planning,
Manipal University, Manipal, Karnataka 576 104, India.*

**Abstract:** Astronomy arises very early in a civilisation and evolves as the civilisation advances. It is therefore reasonable to assume that a vibrant knowledge of astronomy would have been a feature of a civilisation the size of the Harappan Civilisation. We suggest that structures dedicated to astronomy existed in every major Harappan city. One such city was Dholavira, an important trading port that was located on an island in what is now the Rann of Kutch during the peak of the Harappan Civilisation. We have analysed an unusual structure at Dholavira that includes two circular rooms. Upon assuming strategically-placed holes in their ceilings we examine the internal movement of sunlight within these rooms and suggest that the larger structure of which they formed a part could have functioned as an astronomical observatory.

**Keywords:** Harappan Culture; Dholavira, astronomical observatory

## 1 INTRODUCTION

Harappan civilisation is probably the largest and the most sophisticated of the Bronze Age civilisations in the world (Agrawal, 2007; Joshi, 2008; Possehl, 2009). During its peak period, between 2500 BC to 1900 BC, it covered an area of more than 1.5 million square km and traded over several thousand kilometres to western Asia and the horn of Africa (Wright, 2010). The Civilisation itself was settled along the banks and upper reaches of two major rivers east of the Thar Desert in what is now Pakistan and India (Figure 1).

One of its most interesting features is several large and medium-sized settlements in the present day Gujarat region in what is called the Kutch (Chakrabarti, 2004; Rajesh and Patel, 2007). Studies of the sites in the Kutch region suggest that the Little Rann of Kutch was covered with water with a few scattered islands. Several Harappan settlements have been found along the higher points in this region reinforcing the idea that the sites in Gujarat were used as trading outposts from which the Harappans traded with West Asia. This is further reinforced by the nature of the settlements, ports and industries found in this area. Several of these were urban centres but in addition there were villages, craft centres, camp sites, fortified places etc. (see Ratnagar, 2006).

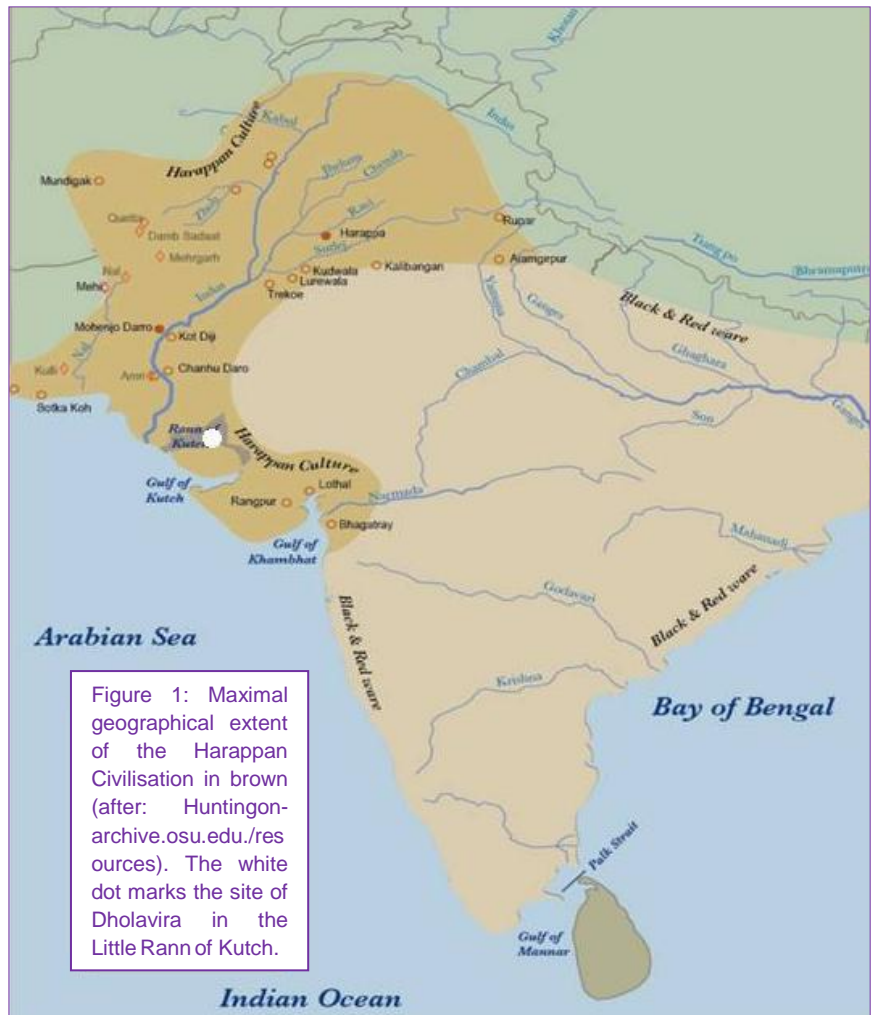

Figure 1: Maximal geographical extent of the Harappan Civilisation in brown (after: Huntingon-archive.osu.edu./resources). The white dot marks the site of Dholavira in the Little Rann of Kutch.

Table 1: Dimensions of Dholavira (after Danino, 2010: 198).

| Location | Measurements (meters) | |
|---|---|---|
|  | Length | Width |
| Lower town (entire city) | 771.1 | 616.9 |
| Middle town | 340.5 | 290.5 |
| Ceremonial ground | 283 | 47.5 |
| Castle (inner) | 114 | 92 |
| Castle (outer) | 151 | 118 |
| Bailey | 120 | 120 |

## 2 DHOLAVIRA

The largest Harappan site in this region of Gujarat was the city of Dholavira (Bisht, 1999), which was on the banks of two seasonal rivulets, and close to a port from which extensive trading is believed to have occurred.

Dholavira was divided into several functional sectors, in keeping with other Harappan cities at this time (Bisht, 2000; Joshi, 2008), and Figure 2 shows a town plan, while dimensions of different parts of the city are listed in Table 1.

For the purpose of this paper, the area of interest is the region referred to as the 'Bailey' in Figure 2, which lies immediately to the west of the Citadel.

### 2.1 The Bailey

In the Bailey region of the city is a structure with a plan-form that is markedly different from all of the other structures in the city and from Harappan plan-forms in general. It consists of the plinth and the foundations of what was probably a 13-room rectangular structure, which included two circular rooms. It is located west of the 'citadel' and is near the edge of the terrace forming the Bailey, which drops off to the west. The flat featureless horizons to the north, west and south are visible without any obstruction, while to the east the mound of the citadel obscures the horizon to a large extent. It is possible that buildings—of which only foundations are visible today—may have obscured the northern horizon to some extent when the city was occupied. The ground slopes down to the south, so it is unlikely that any structures would have obscured the southern horizon (assuming that they were all only single-storied).

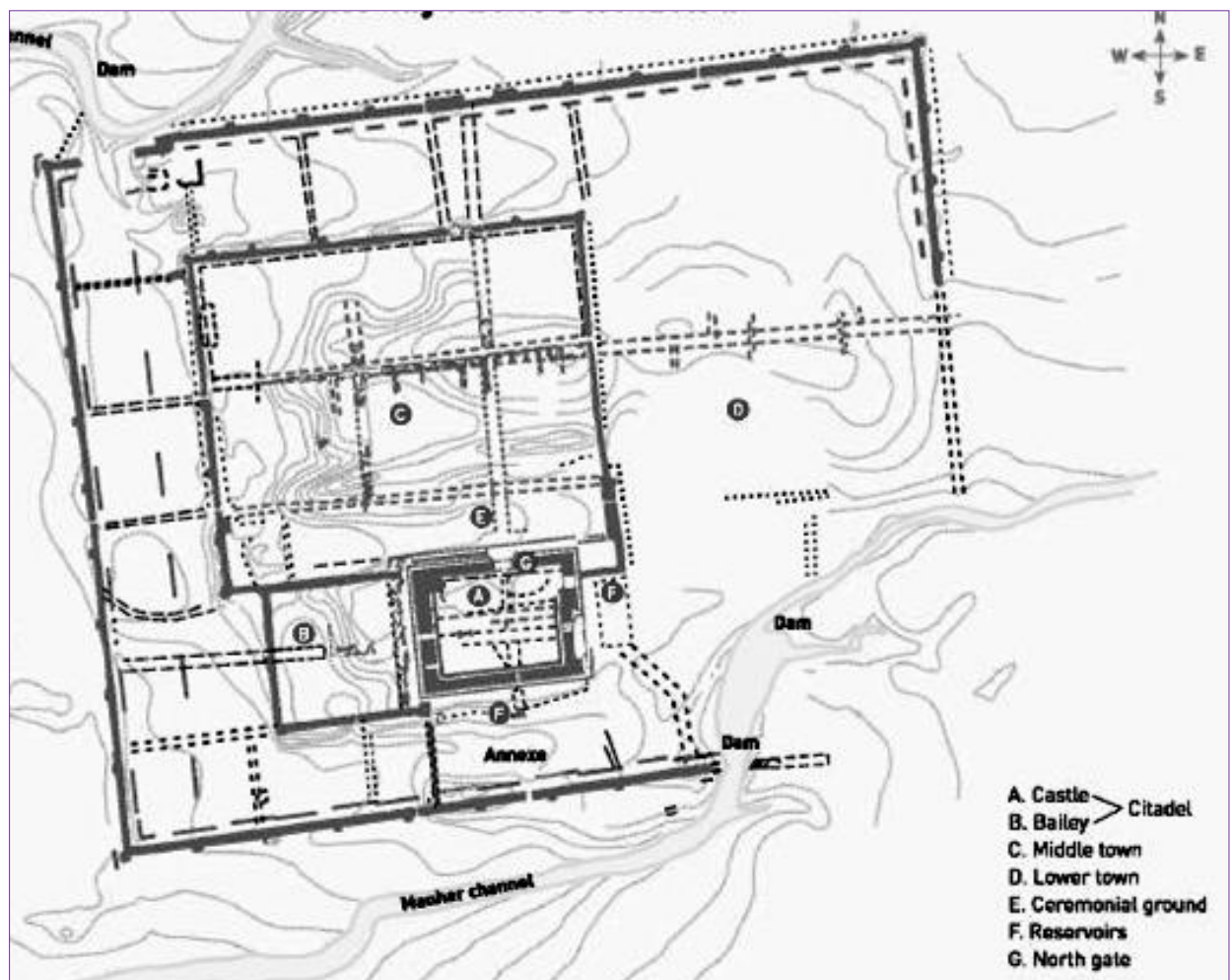

Figure 2: A plan of the site of Dholavira (from the web site of the Archaeological Survey of India, http://www.asi.nic.in/asi_exca_2007_dholavira.asp), showing the location of the 'Bailey' (B). To indicate the scale, the 'Bailey' measures 120 × 120 meters.

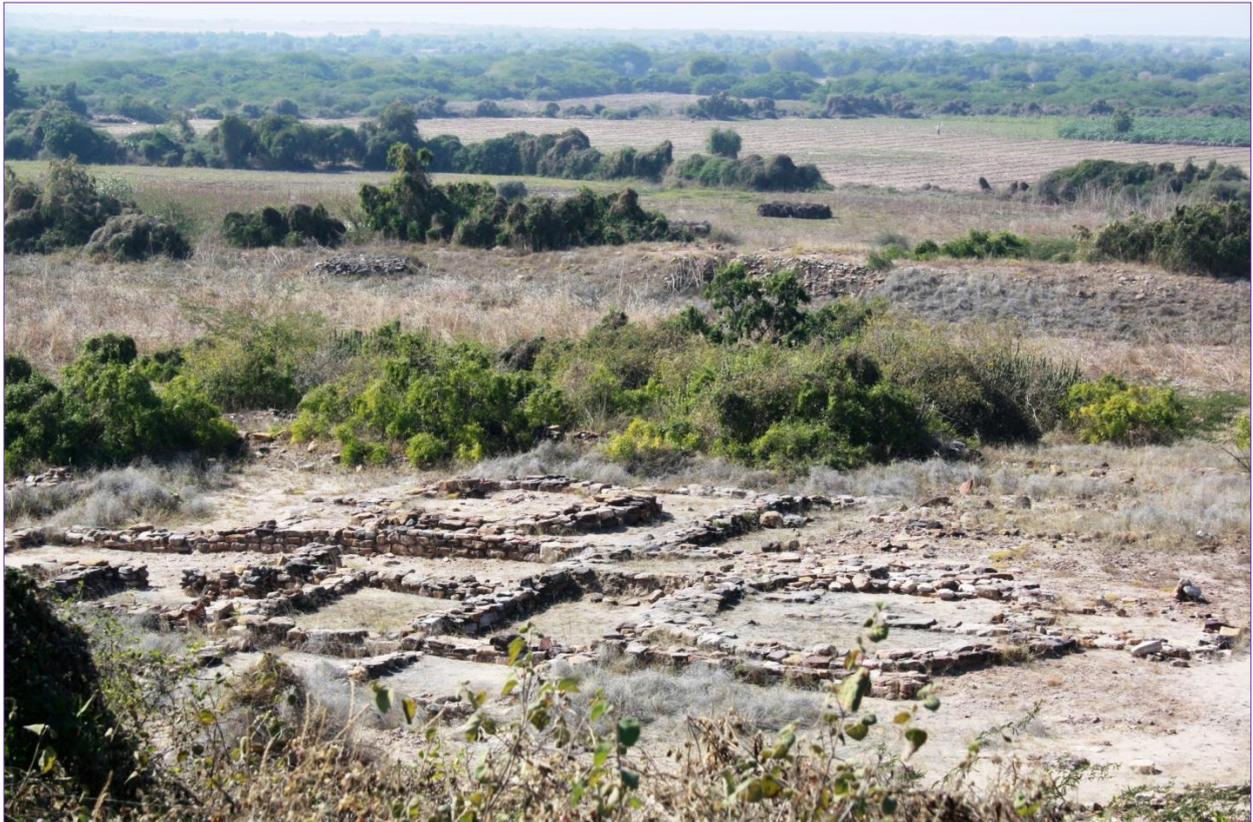

Figure 3: Photograph of the Bailey structure at Dholavira. There are two circular rooms, one to the right of the picture and the other in the centre. The one on the right has a central structure that faces due North.

In Figure 3 is a photograph of the Bailey. As can be seen, the structure consists of rooms of circular and square shape. Since most other residential and workshop buildings in Dholavira are rectangular, it generally has been assumed that the Bailey structure also dates to the Late Harappan Period or an even later Period. However, we suggest that the Bailey structure dates to an earlier period, because of its unique combination of rectangular and circular rooms. Furthermore, we suggest that the two non-circular rooms were designed for non-residential purposes, because:

1) The rectangular rooms adjacent to the circular rooms had bathing and other utilitarian areas, but these were missing from the circular rooms.
2) Each rectangular room typically was connected to one or more other rooms, but each of the circular rooms had only one entrance.
3) Each of the circular rooms was far too small to have served as a residence.

It also is clear that the Bailey structure was built on top of an earlier Harappan structure. We suggest that the entire Bailey area was infilled and reconstructed at the peak period of the city, thereby acquiring its present shape.

### 2.1.1 Survey of the Bailey Structure

In December 2010 we surveyed the remains of the Bailey structure (see Figure 4), and noticed a number of unique features of the construction. Firstly, at three places where E-W oriented cross walls met a N-S oriented wall, they were offset by the thickness of the wall. Since the obvious common sense approach would have been to carry on the cross walls in the same line, this misalignment must have been deliberate (although the reason for this remains obscure).

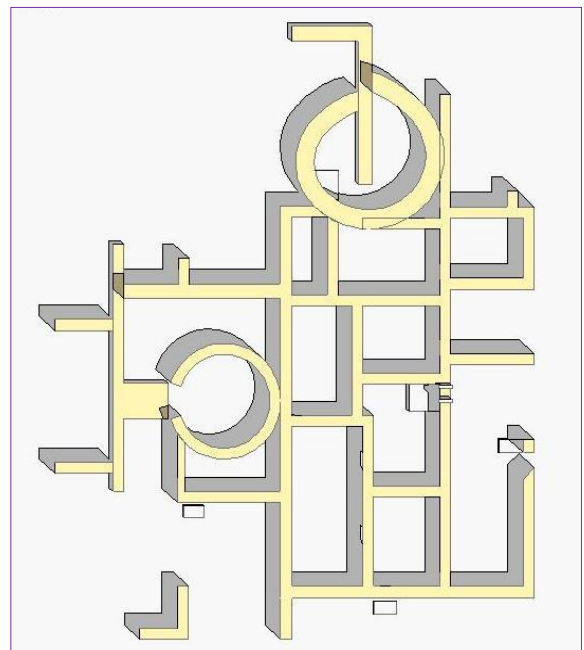

Figure 4: A plan of the Bailey structure at Dholavira showing the two circular rooms. North is to the top.

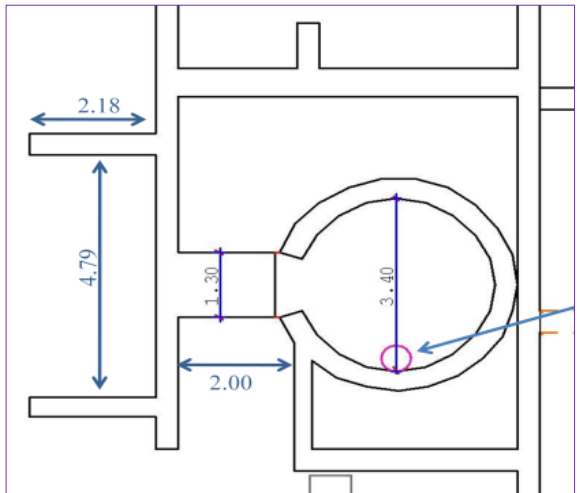

Figure 5: The ground plan and dimensions of the western circular room. All dimensions are in meters. The red circle marked by the blue arrow indicates the location of the presumed hole in the ceiling.

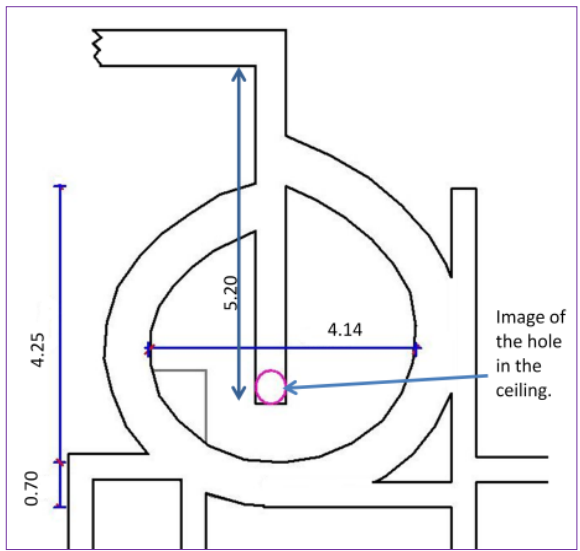

Figure 6: The ground plan and dimensions of the northern circular room. All dimensions are in meters.

As Figure 4 indicates, there are two circular rooms—one in the north and one in the west, and details of these are presented in Figures 5 and 6. The western room is perfectly circular with a mean internal diameter of 3.4 m and a wall thickness of 0.75 m, while the northern circular room is like a spiral in plan such that the line of the outer surface of its wall comes in line with its inner line at its northernmost point as it completes 360° (presumably without letting too much diffused light into the room). A 'straight wall' 0.75 m in thickness extends N-S into the room at this point for 4.0 m. A wedge-shaped segment 1.5 m on two sides and bounded by the curvature of the circular wall of the room is situated in the southwestern quadrant of the room.

In the course of the survey we noticed the following unusual aspects of the entire area:

1) Unlike all other regions of the settlement, the land surface in the Bailey area rises from south to north with an inclination of nearly 23.5°, which corresponds exactly to the latitude of the site. Hence, for someone standing at the southern end of the Bailey, the North Pole would be at the top of the slope, and all stars seen would be circumpolar.
2) While the city of Dholavira is aligned $6 \pm 0.5°$ from true north, features associated with the two circular rooms pointed exactly to the west ($270 \pm 0.5°$; see Figure 5) and the north ($0 \pm 0.5°$; see Figure 6).
3) The circular room at the northern end of the Bailey has a small platform in the southwestern part of the room.
4) At the southern end of the Bailey are two deep rectangular pits (shown in Figure 4). These lack any entry/exit stairs and while their function is not obvious, they possibly could have been used for observing stars at and near the zenith.

Let us now investigate whether the Bailey structure could have had an astronomical function.

### 2.1.2 Our Simulation

Assuming a wall height of 2.5 m (see Bisht, 2000; Danino, 2008; Joshi, 2008) for the Bailey structure and entry to these northern and western circular rooms from the north and west respectively, we simulated the response of the rooms to solar geometry for the latitude of Dholavira. The assumptions and the simulation procedure are detailed below.

For the northern circular room we assumed that the entry point was via a break in the 2.5 m high circular wall where the 'straight wall' penetrates from the north. The width of the entry was taken as 0.50 m—which is the thickness of the 'straight wall'. The 'straight wall' was interpreted as a walkway just 0.60 m high. In keeping with our knowledge of Harappan architecture (ibid.) a flat roof was assumed for the room, with a circular opening 0.50-m in diameter located directly above the termination point of the walkway (see Figure 7).

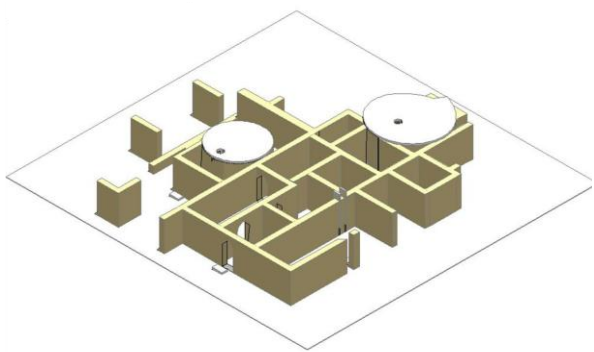

Figure 7: Showing the hypothetical reconstruction of the Dholavira Bailey structure with 2.5m high walls.

Upon simulating the summer solstice day, the circle of light cast by the aperture in the roof slides down the circular wall in the west and across the floor and, at local solar noon, falls directly upon the extreme south portion of the walkway (Figure 8) before continuing across the floor and up the eastern portion of the circular wall. This is expected since we have deliberately positioned the aperture over the southern end of the walkway and the Sun is directly overhead at local solar noon at the time of the summer solstice for the latitude of Dholavira. But what is particularly exciting and caught our attention is that when we simulated the movement of the Sun at the time of the winter solstice, using this same geometry, the circle of sunlight travels down the N-W part of the circular wall and when it is on the top surface of the walkway, its northern edge grazes the bottom edge of the circular wall (see Figure 9).

Similarly, for the western room, we assumed that the entry was via a break in the 2.5 m high circular wall where the straight wall joins from the west. The width of the entry is taken as 1.30 m, which is the thickness of the straight wall. The straight wall is once again taken as a walkway just 0.60-m high. A flat roof was assumed for the structure, with a circular opening 0.50m in diameter at the southern extreme.

Upon simulating for the summer solstice day, the circle of light cast by the aperture in the roof slides down the circular wall in the S-W and is on the floor at local solar noon, its southern edge grazing the bottom edge of the southern wall before continuing up the S-E portion of the circular wall (see Figure 10). This is expected since we have deliberately positioned the aperture over the southern extreme and the Sun is directly overhead at local solar noon on the summer solstice at Dholavira, as mentioned earlier. Simulating the Sun's movement on the day of the winter solstice and using this same geometry, the circle of light travels down the N-W part of the circular wall and when it is on the straight wall, its northern edge passes close to the bottom edge of the circular wall (Figure 11).

In addition, it is seen that the two sections of E-W oriented walls to the west of the west circular room frame the extreme points of the setting Sun as seen from the 1.30 m wide slit in the circular wall. In other words, the shadow of the northern of these walls touches the northern extremity of the slit at sunset on the summer solstice day and that of the southern of these walls touches the southern extremity of the slit on the winter solstice day (Figures 12 and 13).

## 3 DISCUSSION

The city of Dholavira is located on the Tropic of Cancer. Thus the shadows of the Bailey structure

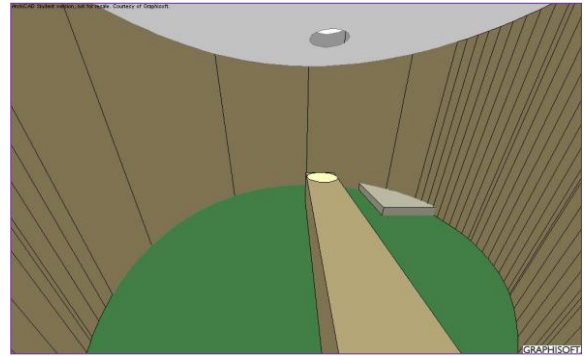
Figure 8: The circle of light cast by the roof aperture for the northern circular room at noon on the summer solstice. The view is from north looking southwards

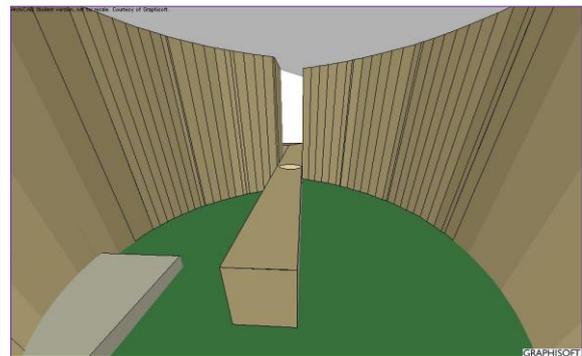
Figure 9: The circle of light cast by the roof aperture for the northern circular room at noon on the winter solstice. The view is from south looking towards the north.

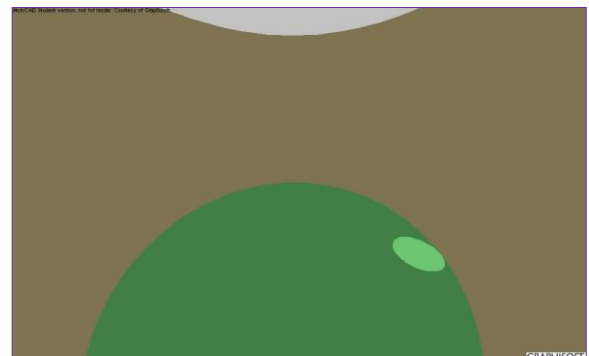
Figure 10: The circle of light cast by the roof aperture for the western circular room at noon on the summer solstice. The view is from west looking towards the east.

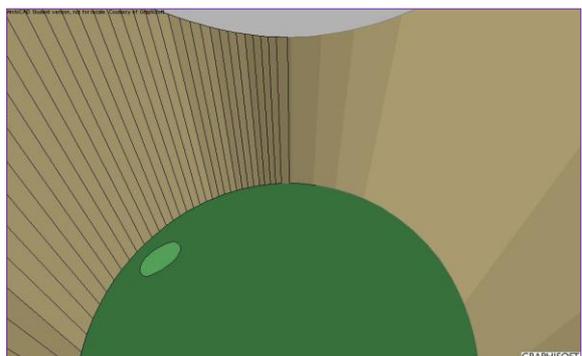
Figure 11: The circle of light cast by the roof aperture for the western circular room at noon on the winter solstice. The view is from west looking towards the east.

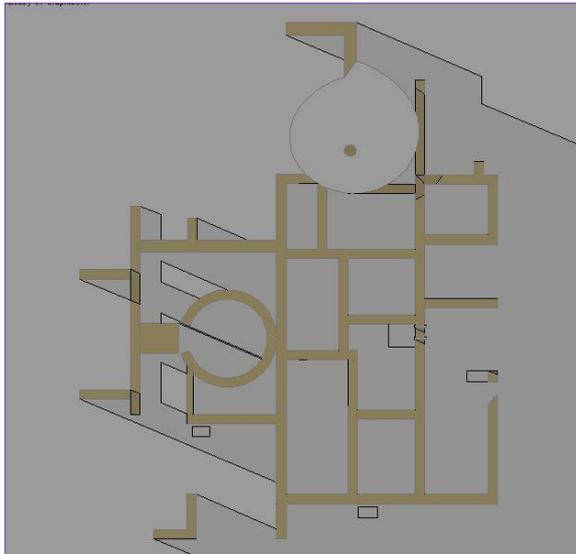

Figure 12: The shadows of the flanking walls (black lines) with respect to the entrance of the western circular room at sunset on the summer solstice.

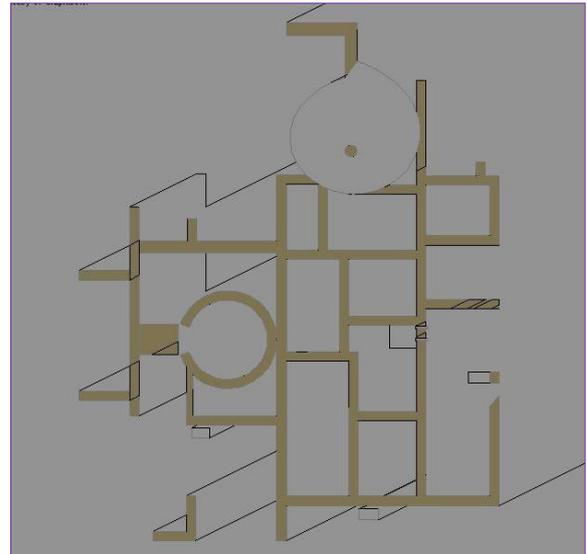

Figure 13: The shadows of the flanking walls (black lines) with respect to the entrance of the western circular room at sunset on the winter solstice.

would always be to the north of that structure, except at noon on the day of the summer solstice when the Sun would be at the zenith and no shadows would be cast. This is clearly something that the Harappan astronomers would have noticed.

The Bailey structure at Dholavira is unusual in several ways. It was built on what seems to be an intentional incline that points to the region of sky where stars always would be circumpolar. The structure also included two circular rooms, a rare occurrence for the 'rectangle-loving' Harappans. However, the workmanship of these two anomalous rooms and their inter-connection with neighbouring ones indicates that they all were contemporaneous. While structures erected by the Harappans normally did not have stone walk-ways leading to their entrances, these two circular rooms had such walkways. While the whole city was inclined 6° to the west of north, the two circular rooms in the Bailey structure had openings that faced due north and due west respectively. In addition, the west-pointing room had two walls to the west that were constructed so that their shadows would just touch the entrance to the room on the winter and summer solstice days.

In seeking to explain the function(s) of the two circular rooms we made assumptions about the superstructure, such as the height of the walls (2.5 meters), the presence flat roofs and the presence of an aperture of a certain size (which was not crucial for our simulations) and positioning. Note that none of these parameters is at variance with what is currently known about Harappan architecture (e.g. see Joshi, 2008; Possehl, 2002).

Adopting these assumptions, we then simulated the movement of the circles of light cast by the holes in the ceilings of these two rooms. In the course of the year these circles of light clearly illuminated specific spots within these rooms, and important days in the solar calendar could easily be identified. The narrow beams of light entering these rooms (which had unusually narrow entrances compared to other rooms at Dholavira) would also have accentuated the movement of the Sun over the course of a year.

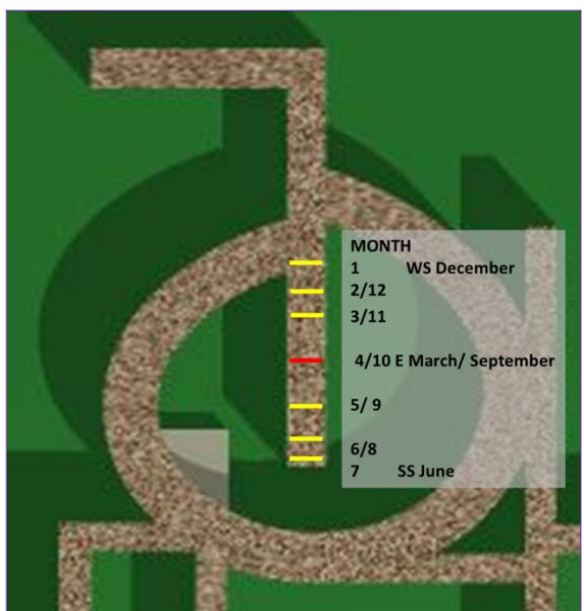

Figure 14: Diagram showing the way in which the northern circular room could act as a calendrical observatory.

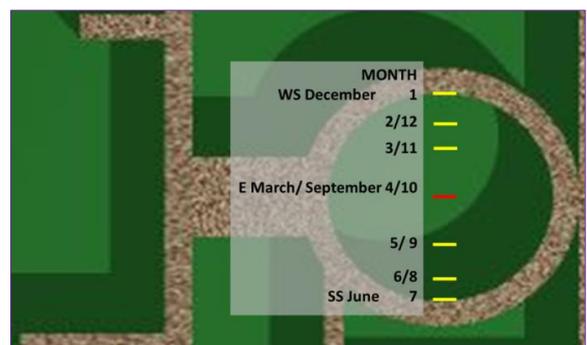

Figure 15: Diagram showing the way in which the western circular room could act as a calendrical observatory.

In the case of the northern circular room, what is interesting is that by positioning the aperture in the roof above the southern extremity of the straight wall, the northern and southern edges of the straight wall marked the points where the circle of light was cast at noon on the solstices. But, in addition, if a marked wooden plank was laid along the walkway, the position of the Sun at noon would vary systematically throughout the year (see Figure 14), and in this way the room also could function as a calendrical observatory.

In the case of the western circular room, once again by positioning the aperture in the roof directly above the southern boundary of the circular wall, the extremes of the N-S diameter of the room marked the points where the circle of light was cast at noon on the solstices. Meanwhile, shadows cast by the two E-W oriented walls to the west of the entrance to the circular room just reached the entrance at sunset on the solstice days, also allowing these specific days to be easily identified. Meanwhile, if a N-S oriented marked wooden plank was laid across the room, this also would show the changing position of the Sun at noon during the year (Figure 15), so this room, too, could serve as a calendrical observatory.

## 4 CONCLUSIONS

It can be safely assumed that astronomers in the intellectually-advanced Harappan Civilization had detailed knowledge of positional astronomy. However, apart from some stray references (e.g. see Maula, 1984; Vahia and Menon, 2011), up to now there has been no positive identification of any structure or artefact with obvious celestial associations at any of the 1500 or so known Harappan archaeological sites.

The Bailey structure at Dholavira is the first Harappan structure that seems to have been constructed specifically in response to the solar geometry at the site, and it is highly probable that the two circular rooms in the structure were designed for solar observations. If this supposition is correct, then this is the first identified Harappan example of a building that was used specifically for observational astronomy. We would argue, however, that similar structures must have existed at all major Harappan cities, and the identification of other examples is simply a matter of time.

Finally, we should mention that since the Dholavira was an important centre of trade and commerce, keeping track of time would have been crucial, but to date no structures that obviously served this purpose have been identified.

## 5 ACKNOWLEDGEMENTS

The authors wish to acknowledge the funding for the project from the Jamsetji Tata Trust under the programme, 'Archaeo-astronomy in Indian Context'. We also wish to gratefully acknowledge the permission given to us by the Archaeological Survey of India to survey Dholavira in 2007, 2008 and 2010. Without this it would have been impossible to carry out our research. We also wish to thank our friends Mr Kishore Menon and others whose endless discussions greatly helped with this work. We wish to thank Professor Vasant Shinde for his continuing encouragement; Professor Sir Arnold Wolfendale for useful suggestions; and Nisha Yadav for her helpful input during this research. Finally, we are particularly grateful to Professor Wayne Orchiston for all of the effort he took to make the contents of this paper precise and clear.

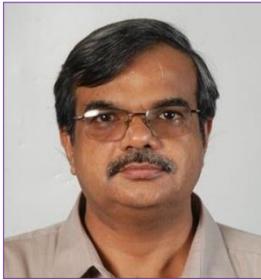

Mayank Vahia has B.Sc. and Master of Physics degrees from the University of Mumbai (India). He is currently a Professor at the Tata Institute of Fundamental Research in Mumbai. He has worked on several projects involving Indian satellites flown on Indian, Russian and American missions to study high energy emission from the Sun and other objects. He has more than 200 publications in most of the major journals in astronomy and astrophysics as well as computer science. Mayank is a member of the IAU Commissions 41 (History of Astronomy) and 44 (Space & High Energy Astrophysics). For the past six years he has been researching the origin and growth of astronomy in the Indian subcontinent and has published about 30 papers on the subject, several of which have appeared in earlier issues of this *Journal*.

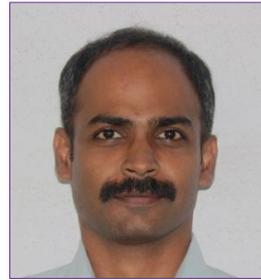

Srikumar M. Menon is an architect with a B. Arch. degree from University of Kerala, India. He has a Ph.D. in archaeoastronomy from Manipal University. He currently teaches in the Faculty of Architecture, Manipal University, India. His research interests are prehistoric architecture of India and early temple architecture of the same region. He is the author of the book *Ancient Stone Riddles: Megaliths of the Indian Bubcontinent* (2013, Manipal, Manipal University Press).